\RequirePackage{fix-cm}
\documentclass[smallextended]{svjour3}       
\smartqed  
\usepackage{graphicx}
\usepackage{multirow}
\usepackage{multicol} 
\usepackage{amssymb}
\usepackage{amsmath}
\usepackage{mathtools}
\usepackage{algorithmic}
\usepackage{url}
\usepackage{graphicx}
\usepackage{hyperref}
\usepackage{color}
\usepackage[top=2cm, bottom=2cm, left=2cm, right=2cm]{geometry}
\usepackage[ruled,vlined]{algorithm2e}
\usepackage{array}
\usepackage{comment}
\usepackage{slashbox}
\usepackage{natbib}

\newcolumntype{H}{>{\setbox0=\hbox\bgroup}c<{\egroup}@{}}

\usepackage{etoolbox}
\apptocmd{\thebibliography}{\renewcommand{\sc}{}}{}{}
\begin{document}

\title{A Hinge-Loss based Codebook Transfer for Cross-Domain Recommendation with Nonoverlapping Data}


\author{Sowmini Devi Veeramachaneni  \and Arun K Pujari \and Vineet Padmanabhan  \and Vikas Kumar}

\institute{Sowmini Devi Veeramachaneni, Vineet Padmanabhan \at
              School of Computer and Information Sciences, University of Hyderabad, Hyderabad, India \\
              Tel.: +91-8985061848\\
              \email{sowmiveeramachaneni@gmail.com, vineetcs@uohyd.ernet.in}           
           \and
           Arun K Pujari \at
              Mahindra University Ecole Centrale School of Engineering (MEC), Hyderabad, India\\
              \email{arun.k.pujari@gmail.com}
              \and
              Vikas Kumar \at
               University of Delhi, Delhi, India\\
               \email{vikas007bca@gmail.com}
}

\date{Received: date / Accepted: date}

\maketitle

\begin{abstract}
Recommender systems(RS), especially collaborative filtering(CF) based RS, has been playing an important role in many e-commerce applications. As the information being searched over the internet is rapidly increasing, users often face the difficulty of finding items of his/her own interest and RS often provides help in such tasks. Recent studies show that, as the item space increases, and the number of items rated by the users become very less, issues like sparsity arise. To mitigate the sparsity problem, transfer learning techniques are being used wherein the data from dense domain(source) is considered in order to predict the missing entries in the sparse domain(target). In this paper, we propose a transfer learning approach for cross-domain recommendation when both domains have no overlap of users and items. In our approach the transferring of knowledge from source to target domain is done in a novel way. We make use of co-clustering technique to obtain the codebook (cluster-level rating pattern) of source domain. By making use of hinge loss function we transfer the learnt codebook of the source domain to target. The use of hinge loss as a loss function is novel and has not been tried before in transfer learning. We demonstrate that our technique improves the approximation of the target matrix on benchmark datasets.


\end{abstract}

\section{Introduction}
Recent years have witnessed a clear explosion in the amount of e-commerce data available. As a result, the quantity of information that needs to be searched for suggesting items of interest to users is becoming more challenging. Recommender systems \cite{10.5555/2931100}\cite{BOBADILLA2013109}\cite{10.5555/1941884} can be used for predicting items that the users will be interested in and has become a core part of several e-commerce applications. Machine learning techniques \cite{article} have become an integral part in the process of building  recommender systems and in the last decade or so there has been a surge in this area of research.
In general, the existing algorithms for recommender systems can be broadly categorized into three major categories namely content-based (CB), collaborative (CF) and hybrid recommendation. In content-based method, the recommendation is made by first calculating a similarity score between a user profile and the profile information of an item. Based on the similarity score, a set of top $k$ items are then recommended to a user. For instance, in movie recommendation task, the profile information of a user contains his/her interest over the \textit{genre}. A movie is then can be recommended by finding out the similarity between the user interest for the \textit{genre} and the movie description (genre). 
In the case of collaborative filtering, the preferences of the target user are matched with that of other users having similar tastes and a recommendation is made.
For example, if a set of users likes the same movies as that of user $A$ (say target user), then there is a chance that user $A$ will like the movies liked by the set of users which the user $A$ has not seen. The hybrid method combines both the content-based and collaborative-based models. 

The major shortcoming with content based recommendation is the situations where the item features are not meaningful or situations where there is a need to capture the change in user interests over time. On the other hand, collaborative filtering handles the above mentioned situations as it only requires the preference (implicit or explicit) information for recommendation. 
The rating matrix is memorised in the memory-based methods and the
missing entries are predicted based on the relation (ratings) between similar users/items. Among memory-based methods, neighbourhood-based methods are prominent wherein users/items that are similar to target user/item are considered for prediction. The idea is that any two users having similar ratings for few items will also tend to have similarity in ratings for the remaining items too. In addition to the assumption, it is also assumed that if rating similarity is there for two items among some users then other users will also give the same rating to the items. Identifying similar users as that of the target user and using the mean of their ratings for prediction is the crux of \emph{user-based} CF methods. In the case of \emph{item-based} CF it is the other way around. Depending on how the weighted average is computed (Pearson correlation, Vector Cosine, Mean-squared-difference etc.), neighbourhood methods may vary.  

In the case of model based methods, missing ratings can be predicted more quickly by making use of a parametric model which is built from the training data.
Low-rank matrix factorisation, which is a latent factor model, is a successful and an effective model-based CF approach. In the Matrix Factorisation (MF) approach as proposed in~\cite{koren2009matrix} and one of the most  widely-used CF technique, a small set of user and item latent factors are learnt based on the user-item rating matrix. By making use of the matrix factorisation method as outlined in~\cite{koren2009matrix} for each user (or, each item) one latent vector is constructed by extracting the latent factors of users/items and this vector captures the user's (or, item's) characteristics. 
To obtain the user ratings for an item, the product of a user latent vector with that of an item is made use through a process called low-dimensional embedding. 
Maximum Margin Matrix Factorisation (MMMF) \cite{srebro2004maximum} is perhaps the most widely used and successful matrix factorisation technique.

Recommender systems often face with the problem of \emph{sparsity} which results from less amount of information being available of the data in hand. 
Machine learning techniques like transfer 
learning~\cite{pan2010survey} has been applied in collaborative filtering based recommender systems to avoid the problems related to sparsity. 
Transfer learning involves the construction of a predictive model across multiple domains wherein relevant knowledge from one domain is extracted and transferred to another domain. Transfer learning in collaborative filtering-based recommender systems is otherwise called as cross-domain collaborative filtering. 
When multiple domains are involved it is necessary for transfer learning to address crucial issues related to the presence/absence  of aligned users/items during knowledge transfer. It is often the case that the presence of common users/items across domains do not exist and even if it does exist it is difficult to map their correspondence. Transfer learning in the absence of common users/items in the two domains is often hard to comprehend and  \emph{CodeBook transfer} (CBT)~\cite{li2009can} has been proposed as a transfer learning approach to address these type of problems. Codebook construction is based on the idea that the abundance of information in one domain is to be made use to predict the missing information in another domain. Since it is assumed that certain  kind of rating pattern is involved in any kind of data, the codebook tries to capture this rating pattern in a summarised form. The belief/hypothesis is that across domains these rating patterns are unvarying.  

Not only in recommender systems, but also in many machine learning algorithms, loss functions plays a significant role in empirical risk minimization and computational complexities \cite{loss1,loss2}. So, choice of loss function is very important.
Among all the loss functions, hinge loss is more suitable for discrete classification task. 
Also, among all the matrix factorisation techniques, maximum margin matrix factorisation(MMMF) treats the collaborative prediction problem as a classification task and takes full advantage of the hinge loss. Until now MMMF has been used in single domain recommender systems and in our proposed approach, we adapt it to work for cross domain recommendation by taking advantage of the hinge loss function.

This paper proposes a codebook based transfer method for collaborative filtering in cross domain recommender systems. Our proposed approach consists of two steps: in the first, we construct the codebook by making use of the source data and apply a technique called co-clustering on it. In the second step, the idea is to transfer the learned codebook in a novel way to the target domain by making use of hinge loss as the loss function. In previous works, researchers have made use of squared loss while transferring the source knowledge to the target domain in order to predict the missing ratings of target domain.  
The proposed codebook based transfer method for collaborative filtering outperforms MMMF (i.e., when MMMF is directly used on target data) and other major codebook based transfer learning approaches for CF as can be seen from the experimental results.

The remainder of the paper is structured in the following manner: Section \ref{rel} presents a quick review of current work on transfer learning in recommendation systems and the proposed approach is explained in Section \ref{prop}. The experimental results are given in Section \ref{expt}, and Section \ref{conclusion} provides the conclusion of the work.
\section{Related work}\label{rel}
Matrix Factorisation has been very popular and successful in predicting the missing values of the user-item rating matrix even when the data is too sparse. The advantage of matrix factorisation technique is that for a matrix which is very sparse and has less information available, the matrix can be filled up accurately to a satisfactory level. Although this is quite good as far as MF techniques for recommender systems are concerned, the major drawback with having more sparsity is the failure in the  accuracy of the prediction levels. To overcome such scenarios researchers have come up with the idea of using information from diverse domains so that information of one domain could be utilized in another. 
Transfer learning \cite{pan2010survey, pan2016survey} is one such technique proposed, which makes use of cross domain collaborative filtering in the domain of recommendation systems to alleviate the sparsity problem. 
Transfer learning intents to transfer the information(knowledge) from a rich(dense) source domain to the sparse target domain. For example, suppose that a particular user has watched many movies and have rated the same. The same user has very less ratings in another domain related to books but wants a book to be recommended, then by using his ratings from the movie domain book can be recommended. Formally, given a dense source rating matrix (source domain), and a sparse target rating matrix (target domain), the goal is to predict the missing entries of target domain by utilizing the information available in the source domain.

The major question that was unanswered is to determine whether both domains are suitable for knowledge transfer. This is necessary because transfer of knowledge can take place only if a correspondence can be established among two domains. Certain assumptions need to be made for applying transfer learning strategy which includes having either common subset of items/users or attribute similarity in items/users.
Domains can be linked and the transfer can happen explicitly via inter-domain similarities, common item attributes, etc. The transfer can happen also implicitly via shared user latent features or item latent features or by rating patterns which can be transferred between the domains.

In~\cite{Chung:2007:IPR:1282100.1282113}, a framework was proposed in which the pertinent items in the source domain are picked based on their common attributes with the target domain (user interested domain). The inter domain links were created utilising the common item attributes, however there is no overlap of users/items required between the domains.
On the other hand the transfer of knowledge by the shared latent
features (of users/items) is addressed in \cite{pan2010transfer}. In this work the latent features of users and items of source domain are learnt and fitted to a target domain by integrating the features into the factorisation of target rating matrix via regularization. However, it require either common users or items among the two domains.
In~\cite{Pan:2011:TLP:2283696.2283784}, the latent characteristics(features) of source and target are shared in a collective way. In this paper a method called matrix factorisation is employed wherein features of both domains are learned simultaneously rather than learning from source first and later applying in the target domain. The only constrain is that in both domains the users and items needs to be identical. The main assumption that is made in this method is that the change of domain does not affect the rating behaviour of the same user/item or different users/items with similar attributes.
In order to maximise knowledge transfer, in~\cite{zhao2013active, ZHAO201738} the authors presume that across different domains there is a correspondence between users/items. Here rather than starting from source domain, matrix factorisation technique is applied first in the target rating data so as to learn the latent features of the target. After that based on certain criteria, random users/items gets selected from the target domain and is matched with the corresponding users/items from source domain. In order to ensure that the latent factors of the selected users/items of both the target and source domain are same, the authors make use of some kind of  regularisation function. 

The methods as outlined in~\cite{itcf, inproceedings, PAN201684} takes care of scenarios wherein the feedback data of target and source are of different types. The users and items in the target and auxiliary(source) domains are assumed to be the same in~\cite{itcf}. 
Two sets of source data are taken into consideration in~\cite{tbt}, one of which shares a common set of users with the target data and the other of which shares a common set of items with the target data. The source data's latent factors are extracted, and similarity graphs are constructed from these latent factors. Both latent factors and similairty graphs are then transferred to the target data.

In collaborative filtering approach that is based on codebook transfer(CBT)~\cite{li2009can} the idea is to capture the hidden correlations that exist between the ratings given by groups of users and groups of items. This pattern is usually denoted by the name codebook or cluster-level rating pattern. In codebook transfer(CBT) what happens is that initially, the codebook of the source domain is extracted by analysing the rating matrix and thereafter this codebook is used for the prediction of the target domain. Regardless of the domains, the way the users and items cluster together it so happens that across domains the cluster level rating behaviour remains invariant.  
\begin{figure*}
\centering
\includegraphics[scale=0.56]{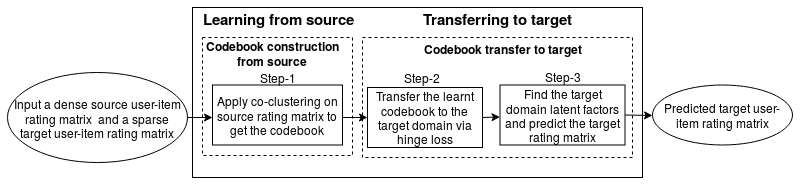}
\caption{Illustration of the proposed method}
\label{blk_diag}
\end{figure*}
One assumption that is made in this connection is regarding the source matrix being a full matrix. Cases wherein the rating matrix is not full, the average rating of a particular user is used for filling-in missing entries of that user.  
Thereafter the rows (users) and columns (items) of the filled-in rating matrix are clustered using co-clustering~\cite{li2009can}. 
The way the codebook is generated is such that each user group is assigned a row and each item group has a column, and each user group and item group has a rating. The average rating of users and items in a particular group is taken into consideration for giving a rating for user-group/item-group. In order to use the codebook in the targt domain it is necessary to select the users/items in the target domain that are more appropriate with the user-group/item-group of the transferred codebook. 

 A linear combination of codebooks from multiple domains is shown  in~\cite{Moreno:2012:TTL:2396761.2396817} wherein using an optimisation technique the coefficient of linear combinations is learned. The assumption of having a fully dense rating matrix is relaxed as demonstrated in~\cite{He:2018:RTL:3159652.3159675, 8233662}. 

A different way of generating the codebook has been outlined in~\cite{ji2016improving, DBLP:journals/asc/VeeramachaneniP19}.
In this method, by making use of the technique of matrix factorisation the user and item latent factors are generated from the source domain and thereby tends to avoid any pre-processing and co-clustering of the rating matrix. The user and item latent factors thus generated are used to obtain the user latent facor group and item latent factor group. Codebook is generated by multiplying the mean latent vecores of the group. 

In~\cite{li2009can}, the authors focus on extracting the group level behavior of users on items by assuming that, though the users/items are different across systems, the groups (groups - based on age, interest etc) of them behave similarly. All the missing ratings in the source rating matrix are filled in a priori by a preprocessing step and then co-clustering is applied on a filled-in source rating matrix to directly get the cluster-level rating pattern (codebook).  The cluster-level rating pattern is then transferred to another domain (target). In the target domain, the user and item membership to clusters encode in the codebook so that a user (or, an item) is member of one user-cluster (or, item-cluster) represented in the codebook. 
In our approach, we do not use a separate preprocessing stage in the source domain, and we use hinge loss instead of squared loss when transferring the learned codebook from the source to the target domain. 
\section{Proposed Method}\label{prop}
Given a dense source user-item rating matrix $X \in \mathbb{R}^{m\times n}$ and a sparse target user-item rating matrix $Y \in \mathbb{R}^{m'\times n'}$ where $m$, $m'$ are the number of users in source and target domain, and $n$, $n'$ are the number of items in source and target data, the goal is to predict the missing entries (as numerous users do not rate many items) in target domain by leveraging the data from the source domain. Prediction of missing entries should take place in such a way that the existing ratings must be approximated with less error rate.
The illustration of the proposed method is shown in Figure~\ref{blk_diag}. Initially (Step-1), the users (rows) and items (columns) of the source rating matrix need to be simultaneously clustered (co-clustering) to construct the codebook.
 We need the cluster indicators of users and items, and in order to get the cluster indicators we can choose any of the co-clustering algorithms. We employed the Orthogonal nonnegative matrix tri-factorisation technique (ONMTF) \cite{Ding:2006} to get the cluster indicators, which has been shown to be equivalent to a two-way k-means clustering approach. The source rating matrix $X$ can be tri-factorised as follows.
\begin{equation}\label{cbfind}
\min_{P, Q, S \geq0}||[X-PSQ^T]\odot W||_{F}^{2}+\alpha||P\textbf{1}-1||_{F}^{2}+\beta||Q\textbf{1}-1||_{F}^{2}
\end{equation}
where $W$ is an indicator matrix and the size of $W$ is $m\times n$. The entry of $W$ is $1$ if the rating exists in $X$ and $0$ for the rest. 
$||.||_F$ represents the Frobenius norm. The dimension of $P$ is $m\times k_1$, $S$ is $k_1\times k_2$, and $Q$ is $n\times k_2$. $P\textbf{1}-1$ ensures the row sum of $P$ to be one, similarly with $Q$. Maximum value of each row of $P$ and $Q$ becomes the cluster indicator for the user/item in that row.
These $P$ and $Q$ are not in a recognizable form in terms of user/item membership matrices. To represent $P$ and $Q$, we use binary values by setting the maximum
valued entry in each row to be $1$ and the others to be $0$. As a result, these binary user/item cluster indicators form (membership) matrices,
denoted by $P_s$ and $Q_s$, for the source rating matrix. From these matrices, we can construct the codebook $B$ as follows.
\begin{equation}\label{cb}
B = [P^T_sXQ_s]\oslash[P^T_s11^TQ_s]
\end{equation}
$\oslash$ indicates element-wise division. Averaging of all the ratings in each of the user/item 
co-cluster takes place in Eq.~\ref{cb}

Once the codebook from the source domain is constructed, transfer (Step-2) it to the target domain ($Y$) by substituting it in Eq.~\ref{tgtmmmf} and solve the objective function ($\mathcal{J}$) in order to get $U$ and $V$ of target data. We use Hinge loss to learn $U$ and $V$ instead of squared loss. As the data of ours is discrete, apart from $U$ and $V$, $r-1$ thresholds
$\theta_{ic}$ $(1\leq c \leq r-1)$ for each user $i$ has to be learned to classify the prediction value into $r$ discrete values. Here, $r$ is the maximum rating (say 5).

\begin{equation}\label{tgtmmmf}
\mathcal{J}(U,V,\Theta) = \sum_{(i,j) \in \Omega}\sum_{c=1}^{r-1}h(\mathcal{T}_{ij}^c(\theta_{ic} - UBV^T)) + \frac{\lambda}{2}(||U||_F^2+||V||_F^2)
\end{equation}
where,
\begin{equation*}
\mathcal{T}_{ij}^c = \begin{cases}
-1 & \text{if $c < y_{ij}$}\\
+1 & \text{if $c \geq y_{ij}$}
\end{cases}
\end{equation*}
$h(\cdot)$ is a smoothed hinge-loss function defined as,
\begin{equation*}
h(d) = \begin{cases}
0 & \text{if $d$ $\geq$ 1} \\
\frac{1}{2}(1 - d)^{2} & \text{if $0 < d < 1$}\\
\frac{1}{2} - d & \text{otherwise.}
\end{cases}
\end{equation*}

$\Omega$ is the set of observed entries, $\lambda >0$ is regularization parameter. 
Gradient based approach can be used to solve the optimization function (Eq. \ref{tgtmmmf}) and following are the gradients.
\begin{equation*}
\frac{\partial\mathcal{J}}{\partial{U_{ik_{1}}}} = \lambda U_{ik_{1}}-\sum_{c=1}^{r-1}\sum_{j|ij \in \Omega}\mathcal{T}_{ij}^c.h'(\mathcal{T}_{ij}^c(\theta_{ic} - UBV^T))VB^T
\end{equation*}
\begin{equation*}
\frac{\partial\mathcal{J}}{\partial{V_{jk_{2}}}} = \lambda V_{jk_{2}}-\sum_{c=1}^{r-1}\sum_{i|ij \in \Omega}\mathcal{T}_{ij}^c.h'(\mathcal{T}_{ij}^c(\theta_{ic} - UBV^T))UB
\end{equation*}
\begin{equation*}
\frac{\partial\mathcal{J}}{\partial{\Theta_{ir}}} = \sum_{j|ij \in \Omega}\mathcal{T}_{ij}^c.h'(\mathcal{T}_{ij}^c(\theta_{ic} - UBV^T))
\end{equation*}
where,
\begin{equation*}
h'(d) = \begin{cases}
0 & \text{if $d$ $\geq$ 1} \\
d-1 & \text{if $0 < d < 1$}\\
-1 & \text{otherwise.}
\end{cases}
\end{equation*}
In gradient descent algorithms, we start with random $U$, $V$, $\Theta$ and iteratively update them using the following update rules.
\begin{equation*}
U_{ik_{1}}^{t+1} = U_{ik_{1}}^{t}-c\frac{\partial\mathcal{J}}{\partial{U_{ik_{1}}^t}}
\end{equation*}
\begin{equation*}
V_{jk_{2}}^{t+1} = V_{jk_{2}}^{t}-c\frac{\partial\mathcal{J}}{\partial{V_{jk_{2}}^t}}
\end{equation*}
\begin{equation*}
\Theta_{ia}^{t+1} = \Theta_{ia}^{t}-c\frac{\partial\mathcal{J}}{\partial{\Theta_{ia}^t}}
\end{equation*}
$c$ is the trade-off parameter.

Once $U$, $V$ of target data are learnt, product of $U$, $B$, and $V^T$ (i.e., $UBV^T$) mapped with the threshold matrix ($\Theta$) becomes the approximation (prediction$\in\{1,2,...,r\}$ say r = 5) of the target rating matrix (Step-3). This is illustrated in Algorithm \ref{algo_prop}.
\begin{algorithm}[ht]
\caption{Codebook transfer via hinge loss} 
\label{algo_prop}
\begin{algorithmic}[1]
\STATE \textbf{Input:}\hspace{10pt} A source rating matrix $X$ of size $ m \times n$ and a sparse target rating matrix $Y$ of size $m' \times n'$ with $y_{ij}$ known for $(i,j)\in\Omega$
\STATE \textbf{Output:}\hspace{2pt} $\hat Y$ ($y_{ij}$ for $(i,j)\notin\Omega$)  
\STATE Tri-factorise $X$ using Eq. \ref{cbfind} inorder to get $P$, $Q$ and $S$.
\STATE Calculate codebook $B$ using Eq. \ref{cb}.
\STATE Transfer the codebook to the target data by substituting in Eq. \ref{tgtmmmf}.
\STATE Solve the optimization (Eq. \ref{tgtmmmf}) using gradient descent technique inorder to get the target $U$, $V$ and $\Theta$.
\STATE Product of $U$, $B$, $V^T$ (i.e., $UBV^T$) by mapping with the threshold matrix gives $\hat{Y}$.
\vspace{-0.1cm}
\end{algorithmic}
\end{algorithm}
\section{Experimental Results and Analysis}\label{expt}
To evaluate the performance of our method we have experimented the method with different datasets. The datasets used are \texttt{MovieLens 100K}\footnote{\label{mvf}https://grouplens.org/datasets/movielens/}, \texttt{MovieLens 1M}\textsuperscript{\ref{mvf}}, \texttt{Goodbooks}\footnote{https://github.com/zygmuntz/goodbooks-10k}, \texttt{Douban Music}\footnote{\label{db}https://github.com/hezi73/TRACER/blob/master/douban.rar}, \texttt{Douban Book}\textsuperscript{\ref{db}}. From the Goodbooks dataset we have considered the first $5000$ users and $3000$ items. From the Douban Music and Douban Book data, we have considered the first $2000$ users and $2000$ items.
The entries of all the datasets fall in \{0,1,2,3,4,5\}, where $0$ indicates the missing value, $1$ indicates that the item is leastly liked and $5$ indicates that the item is heavily liked. The statistics of the datasets are given in Table \ref{dataset}. In all the experiments, we have divided the observed data into 80\% and 20\%, in which 80\% is used for training, and 20\% is used for testing. 
\begin{table*}[ht]
\centering
\caption{Datasets statistics}
\vspace{0.25cm}
\label{dataset}
\resizebox{14.5cm}{!}{
\begin{tabular}{|c|c|c|c|}
\hline
\textbf{Dataset} & \textbf{\# of Users} & \textbf{\# of Items} & \textbf{\% of Observed entries} \\ \hline
MovieLens 100K  & 945  & 1682 & 6.29  \\ \hline
MovieLens 1M    & 6040 & 3952 & 3.77  \\ \hline
Goodbooks (Subset) 	& 5000 & 3000 & 1.08  \\ \hline
Douban Music (Subset) & 2000 & 2000 & 11.26 \\ \hline
Douban Book (Subset) & 2000 & 2000 & 5.51 \\ \hline
\end{tabular}
}
\end{table*}
\subsection{Metrics used for Evaluation}
As far as collaborative filtering algorithms are concerned the computation of the prediction accuracy is crucial for assessing the performance of the algorithms. Root Mean Square Error (RMSE) (Eq. (\ref{rmse})) and Mean Absolute Error (MAE) (Eq. (\ref{mae})) are the two most widely used measures for computing the prediction accuracy. The difference between the predicted rating and the true(actual) rating forms the basis for these measures and it is natural that better performance means smaller values for RMSE and MAE. The values stated in the tables are the average of five runs.
\begin{equation}\label{rmse}
RMSE = \sqrt{\sum\limits_{(i,j)\epsilon \Omega}\frac{{(y_{ij}-\hat y_{ij})}^2}{|\Omega|}}
\end{equation}
\begin{equation}\label{mae}
MAE = {\sum\limits_{(i,j)\epsilon \Omega}\frac{|(y_{ij}-\hat y_{ij})|}{|\Omega|}}
\end{equation}
where $y_{ij}$ is the original rating and $\hat y_{ij}$ is the predicted rating, $|\Omega|$ is the count of test ratings.
\begin{figure}[h]
\centering
\includegraphics[scale=0.8]{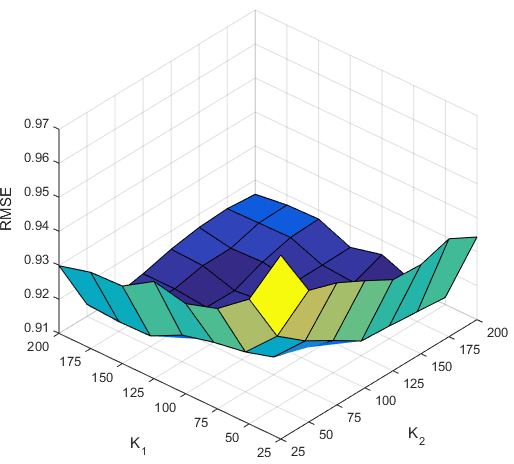}
\caption{Impact of number of clusters ($k_1$, $k_2$) on RMSE of MovieLens 1M data when MovieLens 100K is considered as source}
\label{fig:rmse}
\end{figure}

\begin{figure}[h]
  \centering
  \includegraphics[scale=0.8]{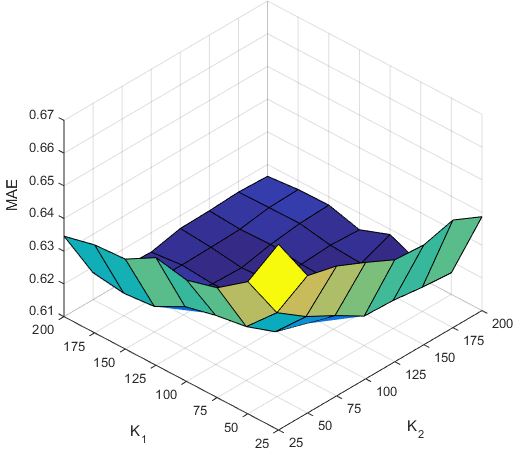}  
\caption{Impact of number of clusters ($k_1$, $k_2$) on MAE of MovieLens 1M data when MovieLens 100K is considered as source}
\label{fig:mae}
\end{figure}

\subsection{Comparison Techniques}
Some of the methods we consider for evaluating our proposed method are outlined below;
\begin{itemize}
\item \textbf{MMMF~\cite{srebro2004maximum,Jas05}:} 
Maximum Margin Matrix Factorisation (MMMF) is the dominant factorisation technique used in collaborative filtering. 
 MMMF is usually applied on the input rating matrix consisting of the user-item ratings. The idea is to find the user and item latent-factor vectors which are of low rank by making use of the existing ratings. 
MMMF can be applied on a single domain only, and hence in our experiments, we applied it on the target domain directly.
\item \textbf{MINDTL \cite{8233662}:} 
In MINDTL, codebook is constructed by taking into consideration the data from all the incomplete source domains. Here codebook for each domain is constructed. 
Following that, the constructed codebooks are linearly integrated and transferred to the target domain, and the absent(missing) values of the target rating matrix gets predicted. As far as our experimental setup is concerned only a single domain is taken into consideration. 

\item \textbf{TRACER \cite{ZHUANG2018287}:} 
In TRACER, data from multiple domains are accounted for and based on this, ratings (which includes missing ratings) for all the source matrices are predicted. Thereafter the predicted knowledge is utilized by transferring it into the target domain. By making use of consensus regularisation during the knowledge transfer process, all the predicted values are forced to be similar. In a way it can be said that in TRACER at the same time learning and transferring happens. 
We have considered a single domain in our experiments, therefore there is no necessity for consensus regularisation. 
We thank the authors for giving the code\footnote{https://github.com/hezi73/TRACER} on the web.
\item \textbf{CBT \cite{li2009can}}: In this approach, the dense part of the source user-item rating matrix is considered, and the missing values of the rows of the dense matrix get imputed using the average of the ratings of particular row (user). The codebook is obtained from the dense user-item matrix by applying the technique of co-clustering. In our experiments, unlike in \cite{li2009can}, which consider only the dense part of the input data, the codebook is constructed by making use of the whole source data and the learnt codebook is transferred to the target domain.

\end{itemize}
\begin{table}[ht!]
\centering
\caption{Values of Root Mean Square Error and Mean Absolute Error of baseline methods and proposed method on MovieLens 1M data, Goodbooks data, Douban Book data}
\label{rmse_cmp1}
\resizebox{8.5cm}{!}{
\begin{tabular}{|c|c|c|c|}
\hline 
\textbf{Dataset} & \textbf{Method} & \textbf{RMSE} & \textbf{MAE}\\ \hline 
\multirow{5}{*}{MovieLens 1M} & MMMF   & 0.9361 & 0.6402  \\ 
\cline{2-4}
&MINDTL  & 0.9948 & 0.7965  \\ 
\cline{2-4}
&TRACER & 0.9800 & 0.8039   \\ 
\cline{2-4}
&CBT & 0.9676 & 0.7746   \\ 
\cline{2-4}
&Proposed & \textbf{0.9123} & \textbf{0.6134} \\ \hline 
\multirow{5}{*}{Goodbooks} & MMMF   & 0.9604 & 0.6524  \\ 
\cline{2-4}
&MINDTL  &  1.2818 & 0.9224  \\ 
\cline{2-4}
&TRACER &  0.9617 & 0.7772  \\ 
\cline{2-4}
&CBT &  0.9634 & 0.7886  \\ 
\cline{2-4}
&Proposed & \textbf{0.9470} & \textbf{0.6381}  \\ \hline 
\multirow{5}{*}{Douban Book} & MMMF   & 0.7976 & 0.5414  \\ 
\cline{2-4}
&MINDTL  & 1.1970  & 0.9054  \\ 
\cline{2-4}
&TRACER &  0.8047 & 0.6600  \\ 
\cline{2-4}
&CBT &  0.7828 & 0.6130  \\ 
\cline{2-4}
&Proposed & \textbf{0.7785} & \textbf{0.5022}  \\ \hline 
\end{tabular}
}
\end{table}
\begin{table}[h]
\centering
\caption{Values of Root Mean Square Error and Mean Absolute Error of baseline methods and proposed method on MovieLens data by considering same dataset as source and target}
\label{rmse_cmp_same}
\resizebox{8.5cm}{!}{
\begin{tabular}{|c|c|c|c|}
\hline
 \textbf{Dataset} & \textbf{Method} & \textbf{RMSE} & \textbf{MAE}\\ \hline
\multirow{5}{*}{MovieLens 100K} & MMMF   & 0.9828 & 0.6808  \\ 
\cline{2-4}
&MINDTL  & 1.9026 & 1.6538  \\ 
\cline{2-4}
&TRACER & 1.0027 & 0.8213   \\ 
\cline{2-4}
&CBT & 1.0264  & 0.8239   \\ 
\cline{2-4}
&Proposed & \textbf{0.9653} & \textbf{0.6600} \\ \hline 
\multirow{5}{*}{MovieLens 1M} & MMMF   & 0.9349 & 0.6389  \\ 
\cline{2-4}
&MINDTL  &  1.8545 & 1.5984  \\ 
\cline{2-4}
&TRACER &  0.9892 & 0.8145  \\ 
\cline{2-4}
&CBT &  1.0729 & 0.8725  \\ 
\cline{2-4}
&Proposed & \textbf{0.9138} & \textbf{0.6175}  \\ \hline
\end{tabular}
}
\end{table}
We have conducted the experiments with varying number of clusters ($k_1$, $k_2$ - 25, 50, 75, 100, 125, 150, 175, 200) and Fig. \ref{fig:rmse} and Fig. \ref{fig:mae} shows the impact of number of clusters on RMSE and MAE of MovieLens 1M data when MovieLens 100K data is considered as source. By observing the figures we can say that the best performance for this scenario (source - MovieLens 100K, target - MovieLens 1M) is achieved when the number of clusters are in between $100$ and $150$ and hence we have fixed both $k_1$, $k_2$ to $125$. In the similar way, when MovieLens 1M is used as source and Goodbooks is the target, the number of clusters is fixed to $150$. Similarly, when Douban music data is used as source and Douban book is the target, we set the number of clusters to $100$.

\par{}Table \ref{rmse_cmp1} shows the values of RMSE and MAE of \texttt{MovieLens 1M} dataset, \texttt{Goodbooks} dataset, \texttt{Douban Book} dataset. For MovieLens 1M data, MovieLens 100K is considered as source and for Goodbooks data, MovieLens 1M is considered as source, and for Douban Book data, Douban music is considered as source data. The first column gives different datasets considered, whereas second column gives the methods considered for comparison along with the proposed method. The third and the fourth columns gives the RMSE and MAE values of the considered methods on datasets. By observing the table, we can say that the proposed method is performing well on any of the datasets considered.

\par{}We have also experimented our method by considering the same datasets as source and target data. It is nothing but reconstructing the same matrix by using the extracted knowledge. Table \ref{rmse_cmp_same} gives the results on \texttt{MovieLens 100k} data when the same is used as source, and also on \texttt{MovieLens 1M} data when itself is used as source. By observing the metric values (RMSE, MAE) in the table, we can claim that our method is outperforming all the comparision methods on the datasets considered.

\section{Conclusions and Future work}\label{conclusion}
A novel method based on codebook based transfer learning for cross domain recommendation has been proposed in this paper. Maximum margin matrix factorisation technique is used to construct the codebook from the source user-item rating matrix.  
By making use of the \emph{hingeloss} function in a novel way, the constructed codebook is transferred to the target domain. The experimental results show that the sparse target matrix is approximated well by the proposed method. Rather than considering only ratings of items, in the future it would be worthwhile to consider social tags as well as other types of data. Applying transfer learning in other real world scenarios other than recommender systems could be another direction in which this research could be taken forward.



%
%


\bibliographystyle{acm}
\bibliography{cbt_prop}

\end{document}